\documentclass[prb,twocolumn,showpacs,floatfix,amsfonts]{revtex4}

\usepackage{graphicx,graphics,color,epsfig}

\usepackage{bm}

\usepackage{amsmath}

\usepackage{amssymb}

\begin{document}

\preprint{}

\title{ Quantum interference in  dirty $d$-wave
superconductors }
\author{Hong-Yi Chen$^*$, Lingyin Zhu$^\dagger$, and C.S. Ting$^*$}
\affiliation{$^*$Texas Center for Superconductivity and Department of
Physics, University of Houston, Houston, TX 77204 \\
$^\dagger$Department of Physics, State University of New York at Buffalo,
Buffalo, NY, 14260}

\begin{abstract}
The local differential tunneling conductance on a Zn impurity in a
disordered $d$-wave superconductors is studied.  Quantum
interference between many impurities leads to definitive
quasiparticle spectra. We suggest that an elaborate analysis on
impurity-induced spectra with quantum interference effect included
may be able to pin down the sign and strength of the scattering
potential of a Zn impurity in low density limit. Numerical
simulations calculated with appropriately determined impurity
parameters are in satisfactory agreement with the observations from
scanning tunneling microscopy (STM) experiments even in subtle
details.
\end{abstract}


\maketitle

Impurity effect has served as a unique probe  to the mechanism of
unconventional superconductivity. Recently, remarkable
improvements in high resolution STM experiments provided
unprecedentedly delicate images of electronic structure around
doped Zn in Bi$_2$Sr$_2$CaCu$_{2}$O$_{8+\delta}$ (BSCCO)
\cite{pan2000}. The defect states in differential conductance map
at low temperatures bear on almost identical spectroscopic and
spatial structures from one impurity to the other, with
characteristic on-site zero bias resonance peak, four-fold
symmetry pattern in local density of states (LDOS) and locally
suppressed superconductivity at immediate surroundings of Zn
atoms. These observations are in qualitative agreement with
preceding theoretical works which model Zn atoms as isolated
pointlike unitary centers embedded in a superconducting background
of $d$-wave like pairing symmetry \cite{balatsky1995,
balatsky0411}. However, although the single-impurity scenario (SI)
is approximately successful in interpreting the  energy and
spatial symmetry of resonant states, it fails to reproduce the
subtle features in the spatial profile of LDOS extracted by STM
measurements. Attempts to reconcile this discrepancy include
postulations on the tunneling matrix of STM probes
\cite{zhujx2001, martin2002}, possible Kondo physics from
staggered moment \cite{sachdev} and etc.; nevertheless, the
determination of the attributes of Zn impurities, with which we
can substantiate reliable calculations to compare with
experiments, remains divergent itself: even assuming Zn impurities
are purely potential scatterers, one could still reach completely
opposite conclusions that they are either repulsive (as desired to
yield unitary limit with realistic cuprates band) or attractive
({\it relative to the background Cu ions}) according to their
ionic configuration. Most recently, an {\it ab-initio} calculation
based on density functional theory concludes that Zn should be
repulsive centers to electrons \cite{wangll0505}.

In the meanwhile, the effect of {\it interference} between the
quasiparicle resonances cannot be neglected. Overlapping between
disorder-induced Fridel oscillations alters local landscape in
LDOS and populates the low energy excitations, which greatly
modify the spectroscopic and transport properties of cuprates.
There have been numerous theoretical studies on many-impurity
problem \cite{onishi1996, pepin2001, morr2002, andersen2003,
atkinson2003, zhuly2003}, with emphasis either on the effect of
impurity network on Fourier transformed power spectrum which can
be used to extract the kinematics of pure systems or on the
asymptotic behavior of Fermi level density of quasiparticle
states. Despite the creditable achievements of these endeavor in
interpreting STM data, the lingering unjustification of impurity
parameters could place their arguments on a shaky ground;
moreover, none of them have given full attention to the potential
of combining interference effect to discuss the impurity spectra
and identify the impurity characteristics itself. It is then the
purpose of this paper to argue that through elaborate
investigation on quantum interference effect in a fully disordered
system, we could nail the identities of Zn impurity such as its
sign and strength and hopefully close the debate on this issue to
our best.

We start by writing down the generic many-impurity Hamiltonian as,
\begin{eqnarray}
H_{Zn} = \sum_{{\bf i}\sigma} \phi(r_{\bf i}) c_{{\bf i}\sigma}^{\dagger}
c_{{\bf i}\sigma}^{} \;,
\end{eqnarray}
where $c_{{\bf i}\sigma}^{\dagger}$($c_{{\bf i}\sigma}$) creates
(annihilates) an electron with spin $\sigma$ at site $r_{\bf i}$
and $\phi(r_{\bf i})$ is the impurity strength on the perturbed
site. Randomness of impurity distribution makes counting relative
coordinates information between all the impurities a formidable
task.  Therefore a systematic and reliable manipulation of the
interference, yet not losing the reality and generality, will be
helpful. We tacitly suggest to fix the first impurity at the
center ($r=0$) of the lattice and keep injecting others while
requiring that the inter-impurity distance of any pair, i.e.,
$R_{\bf ij}=|r_{\bf i}-r_{\bf j}|$ has to be greater than a preset
cutoff $d_0$. This constraint excludes the possibility of cluster
formation and further guarantees a uniform distribution of
disordered sites. The central impurity, which is the focus of our
discussion later,  is believed to be the representative of a
arbitrary one surrounded by other disordered sites that are
randomly distributed in real samples.  Therefore, a single
parameter, i.e., the concentration, will be the defining parameter
instead of the relative orientation and distance between
impurities.  The effect of the quantum interference is then
parameterized as a function of the concentration $x$ (or the
number of impurities $N_{imp}$) in a fixed lattice size.

We study this problem  using the  conventional Bogolibuv-de Gennes' (BdG)
formalism:
\begin{eqnarray}
\sum_{\bf j}^N \left(
   \begin{array}{cc}
      {\cal H}_{\bf ij} & \Delta_{\bf ij}^{} \\
      \Delta_{\bf ij}^* & -{\cal H}_{\bf ij}^*
   \end{array} \right) \left(
   \begin{array}{c}
      u_{\bf j}^n \\
      v_{\bf j}^n
   \end{array} \right) = E_n \left(
   \begin{array}{c}
      u_{\bf i}^n \\
      v_{\bf i}^n
   \end{array} \right)\;,
\end{eqnarray}
where
${\cal H}_{\bf ij} =
   - t_{\bf ij}+\left(\phi(r_{\bf i})-\mu\right)\delta_{\bf ij}$
is the single particle Hamiltonian, $t_{\bf ij}$ is the hopping integral,
 $\mu$ is
the chemical potential,
$\Delta_{\bf ij}^{} =
   \frac{V}{2}\langle c_{{\bf i}\uparrow}^{} c_{{\bf j}\downarrow}^{}
   - c_{{\bf i}\downarrow}^{} c_{{\bf j}\uparrow}^{} \rangle$
is the spin-singlet bond order parameter, $V$ is the
nearest-neighbor attractive potential), $u_{\bf i}^n$ and $v_{\bf
i}^n$ are the eigenfunctions of the BdG equations, and $E_n$ is the
eigen-energy. The self-consistent conditions are applied to solve
the BdG equations:
\begin{eqnarray}
& {} \langle n_{{\bf i}\uparrow} \rangle =
   \displaystyle{\sum_{n=1}^{N}} \left| u_{\bf i}^n \right|^2 f(E_n)\;,
   {}& \\
& {} \langle n_{{\bf i}\downarrow} \rangle =
   \displaystyle{\sum_{n=1}^{N}} \left| v_{\bf i}^n \right|^2
   [1-f(E_n)]\;, {}& \\
& {} \Delta_{\bf ij}^{} =
   \displaystyle{\sum_{n=1}^{N}} \frac{V}{4} \left( u_{\bf i}^n v_{\bf
   j}^{n*} + v_{\bf i}^{n*} u_{\bf j}^n \right) \tanh \left( \frac{\beta
   E_n}{2} \right) \;, {}&
\end{eqnarray}
where $f(E)=1\slash(e^{\beta E}+1)$ is the Fermi distribution.  We use
the following parameters throughout this paper: $\langle t_{\bf ij}
\rangle =t =150$ meV, $\langle t_{\bf ij} \rangle =t'=-0.3t$, $V=1t$.
This prepares us a hole-like Fermi surface with the hole doping
$1 - n_f =
   1 - \sum_{{\bf i} \sigma} \langle c_{{\bf i}\sigma}^\dagger c_{{\bf
   i}\sigma}^{} \rangle / N_xN_y = 0.15$,
i.e., optimally doped region and a maximum gap magnitude about 45
meV. The local density of states takes the following expression:
\begin{eqnarray}
\rho_{\bf i}(E) &=& -\frac{2}{M_x M_y} \sum_{n{\bf k}}^{N} \biggl[
\left| u_{\bf i}^{n{\bf
   k}} \right|^2 f'(E_{n,{\bf k}}-E) \nonumber \\
&& +\left| v_{\bf i}^{n{\bf k}} \right|^2 f'(E_{n,{\bf k}} + E)
\biggr],
\end{eqnarray}
where the factor of 2 arises from spin degeneracy.  The summation in
$\rho_{\bf i}(E)$ is averaged over $M_x \times M_y$ (in our paper 20
$\times$ 20) wavevectors in the first Brillouin zone.

{\it Single impurity} We start our discussion with a brief review on
the single impurity model and plot the resulting spectra in Fig.1
with the ``blocking effect''(which is addressed below) included. We
unbiasedly picked up a repulsive potential $\phi=3.0$eV. Because of
the Bogoliubov symmetry of quasiparticles, the on-site LDTC is
expected to display two resonance peaks at $\Omega^{\pm}_0=\pm
0.015t$ \cite{balatsky1995}[see Fig.1(a),(b)]. While the spatial
scale and symmetry of impurity states observed in STM fit well with
the prediction of SI theory, the relative spectral weight
distribution is completely reverted \cite{balatsky0411, zhuly2003}:
experimental images at $\Omega$ = -1.5 meV \cite{pan2000} displays
an on-site maximum spectral intensity, a local minimum on its
nearest neighbor sites and a second maximum on the next nearest
sites, as illustrated schematically in Fig.1(d).
 This inconsistency gave birth to the conjectures
regarding the indirect tunneling through the insulating BiO layer:
the ``filter effect'' due to the coherent tunneling
\cite{martin2002} and an alternative ``blocking effect'' due to the
incoherent tunneling \cite{zhujx2000}.  In our paper, we will stick
to the argument of the latter, in which the ``local differential
tunneling conductance (LDTC)" fetched by STM probes is believed to
measure acutally the averaged LDOS over the four nearest neighbor
sites, i.e.,
\begin{eqnarray}
G_{\bf i}(E) = \frac{1}{4} \sum_{\bf\hat e} \rho_{\bf i+\bf\hat e}(E)\;,
\end{eqnarray}
where $\bf\hat e$ is $\pm \hat{x}$ or $\pm \hat{y}$. Fig
1.(a),(b),(c) are then plotted with the ``blocking effect" included
intentionally. In Fig.1(c), the spatial distribution of spectral
intensity mimics experimental results under this manipulation.
However, corresponding on-site spectra [Fig 1(a)]deviates from the
result of Pan {\it et al.} remarkably: the former shows a sharp peak with
excess spectral background at negative bias and a second resonance
at positive side; the latter displays a zero bias conductance peak
and a second small peak on negative bias; furthermore, Pan {\it et al.}
showed that the Friedel oscillation along the nodal direction decays
surprisingly faster than that of the antinodal direction. This is
rather counterintuitive since the subgap resonances in $d$-wave
superconductors overlap with the vanishing continuum and hence are
virtually bounded, with supposedly extended tails along the nodal
direction due to the vanishing order parameter. We argue that with
the inclusion of quantum interference effect, all this discrepancy
will be reconciled provided that the impurity parameter is properly
settled.

\begin{figure}[!t]
\centerline{\epsfxsize=8.5cm\epsfbox{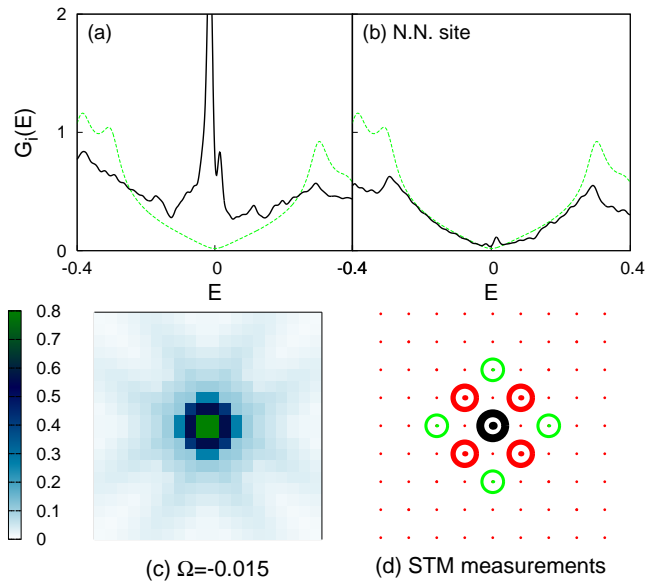}} \caption[*]{(a) LDTC
on the impurity-site. Dashed green: pure system; solid black: LDTC
with the present of single impurity;  (b)  LDTC on the nearest
neighbor site. Dashed green: pure system; solid black: LDTC with the
present of single impurity;  (c) calculation on spatial distribution
of LDTC at $\Omega_0=-0.015t$; impurity site is placed at the center
of the squared window. (d) Schematic plot of the STM image of one
impurity \cite{pan2000}.}
\end{figure}

\begin{figure}[!t]
\centerline{\epsfxsize=8.5cm\epsfbox{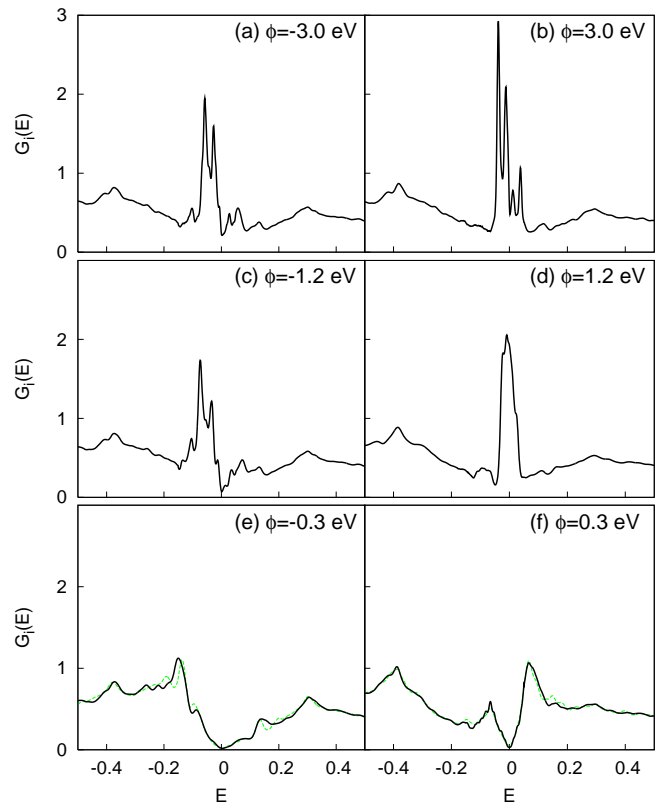}} \caption[*] {The
local differential tunneling conductance (LDTC) on the central Zn
impurity with 10 impurities embedded in the lattice. Left panel:
negative potentials (a) -3.0 eV (c) -1.2 eV (e) -0.3 eV. Right
panel: positive potentials (b) 3.0 eV (d) 1.2 eV (f) 0.3 eV. Green
curves in (e) and (f) are on-site LDTC of single-impurity case.}
\end{figure}

{\it Many impurity} We investigate the quantum interference effect
by embedding 10 impurities in an otherwise clean lattice, which
yields $x=0.41\%$, close to laboratory terms. The lattice size is of
$N_x \times N_y = 49 \times 49$. Fig. 2 enumerates the on-site LDTC
for different values of impurity strength systematically. Generally,
quantum interference splits the single impurity resonances into
multiple peaks, as illustrated in Fig. 2 and extracting distinctive
information from those peaks directly is not easy.
However, the zero energy residual LDOS $\rho(\omega=0)$ is
particularly useful as it evolves systematically with respect to the
impurity potential. When $\phi=-3.0$eV, $\rho(\omega=0)$ is
considerably large; as $\phi$ increases but remains negative ({\it
attractive}), $\rho(\omega=0)$ decreases until it is completely
depleted. This tendency can be understood qualitatively in the sense
that the sample recovers its clean spectra which has vanishing
$\rho(\omega=0)$ when  less contaminated. When $\phi$ becomes
positive and increases  ({\it repulsive}), $\rho(\omega=0)$
aggregates remarkably and multiple peaks merge into a ``single
resonance peak'' when $\phi$ is around the value with which unitary
limit is obtained in single impurity analysis. Fig. 2(d) shows such
a strong resonance peak at negative bias slightly below zero, whose
width and height are in fair agreement with the result of Pan {\it
et al.}. The likeness between Fig. 1(a) and Fig. 2(d) will plausibly
suggests the single-impurity physics as dominating mechanism and the
fact that Zn impurities in the experimental conductance maps appear to be
isolated entities with nearly undisturbed four-fold symmetry in
LDTC seemingly supports this viewpoint.  However, it is worthwhile
to point out that exact width of this zero-bias resonance in Fig.
2(d) and STM experiments \cite{pan2000} is $\sim 10$ meV, an order
of magnitude bigger than what single-impurity theory expects but
agrees well with the impurity bandwidth in unitarity low density
limit, i.e., $\gamma = \sqrt{n_i\Delta_0E_F}$ \cite{sctma}. Hence
the figurative resemblance between the on-site LDTC's of
many-impurity and single-impurity must be the consequence of
quantum interference between dilute impurity states and the
``homogeneous broadening'' effect which was proposed by Atkinson
{\it et al.} \cite{atkinson2003} and is not negligible in
laboratory terms. The possibility of $|\phi|$ being weak (less
than 0.4 eV) is excluded, as it will not introduce any prominent
resonant behavior and observable interference effect requires
inter-impurity distance as close as $3a$, corresponding to an
unreasonably large population of disordered sites. This is
demonstrated in Fig. 2(e) and 2(f): when $\phi=\pm0.3$ eV, the
on-site LDTC of dirty samples are essentially coincident with that
of single-impurity case.  Collection of all the facts above and
comparison with experiments lead us to believe that Zn atoms in
BSCCO-2212 be {\it repulsive } potentials at least in the zeroth
order.

\begin{figure}[!t]
\centerline{\epsfxsize=8.5cm\epsfbox{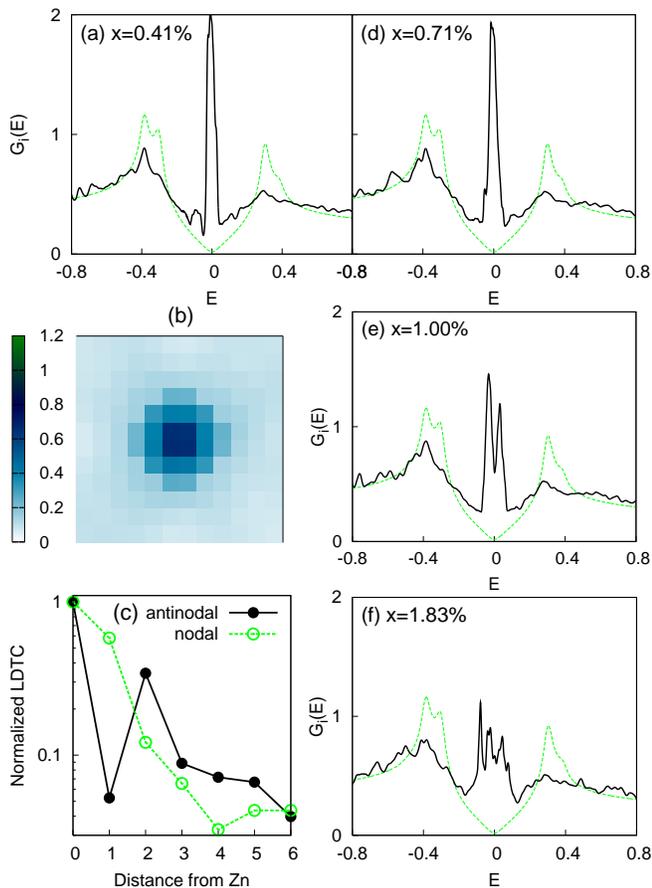}} \caption[*] {(a)
The LDTC $G_{\bf i}(E)$ on a Zn impurity at $r=0$ for $x=0.42\%$.
The dashed line represents the LDTC far away from Zn impurities;
 (b) The spatial profile of LDTC maps at
$\Omega=-2.4$ meV;(c) The LDTC (normalized to the peak value)
versus distance from the Zn impurity;  (d)(e)(f) The LDTC $G_{\bf
i}(E)$ on the Zn impurity at $r=0$ for $x=0.75\%,1.0\%,1.83\%$.}
\end{figure}

 The exact magnitude of Zinc impurity is still open to
determination, but favorably it should not differ too much from
the value which approximates single-impurity unitary limit (which
is $\phi=1.2$eV here) with the cuprates band structure  since the
unitary scattering due to doped Zn is verified by the scattering
phase shift of  $0.49\pi$ in experiments \cite{pan2000}.  We then
adopted $\phi=1.2$eV as the input of  calculations and the results
are plotted in Fig. 3.  The spectrum in channel (a) shows a
striking agreement with the result of Pan {\it et al.} with a sharp
resonant peak forming at $\Omega=0$ and a second smaller peak on
negative bias. Apparently, single-impurity scattering fails to
interpret both the asymmetry of the peaks positions and the
broadened peak widths. While the staggered magnetic interaction
may also address the two-peak structure, we would rather believe
that the negative peak should arise collectively from
inter-impurity correlations and our numerical inspection confirms
that it actually persists in a fairly wide region of positive
$\phi$'s; in Fig. 3(b), the spatial distribution of the resonance
around a Zn impurity shows the local minima of the four
nearest-neighbor sites, and the local maxima of the eight
2nd-nearest- and 3rd-nearest-neighbor sites clearly forming a
``box'' around this Zn atom; in Fig. 3(c), the normalized LDTC
$G({\bf r})$ at resonance frequency is plotted as a function of
distance away from the impurity along the nodal and antinodal
directions. The strength of the normalized LDTC in the nodal
direction is found to decay faster than that along the antinodal
direction,  reinforcing what is observed in STM experiments
\cite{pan2000}. All this above agree with STM observations
remarkably and are confirmed to be robust against the change of
$\phi$ within a wide range centered at $1.2$eV;    When impurity
concentration increases, the low energy excitations are further
populated and a subgap impurity band is established gradually,
which will suppress superconductivity eventually and is referred
to as ``impurity band", in analogy to similar phenomena in
semiconductors. This is illustrated schematically in Fig.
3(d)(e)(f), where the zero bias impurity resonance is further
broadened and finally into multiple-peak structure with excessive
spectral weight filling up the gap region when impurity
concentration is several times bigger than the experimental value
($0.2\%-0.5\%$).

{\it Conclusion}: We discuss how quantum interference between many
impurities can produce qualitatively different spectral features
when the  impurity variables change.  With a numerical
study on the disorder induced local spectra (on impurity site), we
would like to close the debate over the identity of Zn impurity by
concluding that Zn atoms in Bi$_2$Sr$_2$CaCu$_2$O$_{8+\delta}$ are
{\it repulsive} to electrons in nature with a strength close to
unitary limit provided that  other internal degrees of freedom was
not considered. The discrepancy between STM experiments and the
results of single-impurity analysis can be reconciled
satisfactorily by taking the quantum interference and insulating
layer blocking effect into consideration simultaneously. Our
numerical calculations with impurity parameter determined in this
paper match STM experiments up to subtle details. We then would
emphasize that the results obtained in laboratories should
actually be interpreted within the frame of collective quantum
interference processes rather than the single impurity physics.

${\bf Acknowledgements}$: We thank S.H. Pan, J.X. Zhu, J. O$'$Neal,  Ang
Li and P.J. Hirschfeld for stimulating comments and suggestions.  This
work is supported by the Texas Center for Superconductivity at the
University of Houston, and by a grant from the Robert A. Welch Foundation
under No. E-1146.

\end{document}